\documentclass{article}

\usepackage{arxiv}
\usepackage{amsmath}
\usepackage[utf8]{inputenc} 
\usepackage[T1]{fontenc}    
\usepackage{hyperref}       
\usepackage{url}            
\usepackage{booktabs}       
\usepackage{amsfonts}       
\usepackage{nicefrac}       
\usepackage{microtype}      
\usepackage{lipsum}		
\usepackage{graphicx}
\usepackage{natbib}
\usepackage{doi}

\title{Application of Black-Litterman Bayesian in Statistical Arbitrage}

\date{Jan, 2023}	

\author{ \href{https://orcid.org/0009-0000-4552-448X}{\includegraphics[scale=0.06]{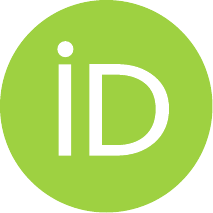}\hspace{1mm}Qiqin Zhou}\thanks{author github: https://github.coecis.cornell.edu/qz247} \\
	Department of Operation Research and Information Engineering\\
	Cornell University\\
	Ithaca, NY \\
	\texttt{qz247@cornell.edu} 
}

\renewcommand{\headeright}{Research}
\renewcommand{\undertitle}{Research}
\renewcommand{\shorttitle}{Research}

\hypersetup{
pdftitle={Application of Black-Litterman Bayesian in Statistical Arbitrage},
pdfsubject={q-bio.NC, q-bio.QM},
pdfauthor={Qiqin Zhou},
pdfkeywords={Bayesian Estimation, Black-Litterman Model, Portfolio Management, Statistical Arbitrage, co-integration, Mean Reversion, Pairs Trading, Quantitative Strategies, Machine Learning},
}

\begin{document}
\maketitle

\begin{abstract}
In this paper, we integrated the statistical arbitrage strategy, pairs trading, into the Black-Litterman model and constructed efficient mean-variance portfolios. Typically, pairs trading underperforms under volatile or distressed market condition because the selected asset pairs fail to revert to equilibrium within the investment horizon. By enhancing this strategy with the Black-Litterman portfolio optimization, we achieved superior performance compared to the S\&P 500 market index under both normal and extreme market conditions. Furthermore, this research presents an innovative idea of incorporating traditional pairs trading strategies into the portfolio optimization framework in a scalable and systematic manner.

\end{abstract}

\keywords{Bayesian Estimation \and Black-Litterman Model \and Portfolio Management \and Statistical Arbitrage \and co-integration \and Mean Reversion \and Pairs Trading \and Quantitative Strategies\and Machine Learning}

\section{Introduction}
We found that integrating pairs trading with the Black-Litterman model portfolio optimization gives us more capibility and flexibility to control for risk and transcation cost. Our backtest shows that the optimized portfolio based on pairs trading strategy significantly enhances the return profile and minimize the down-side risk during financial crisis. 

In this paper, we first developed a traditional pair trading strategy with co-integration modelling to capture asset mispricing. Then we connected pairs trading strategy to a portfolio optimization problem using the Black-Litterman Model. Black-Litterman Model employs the Bayesian approach to estimate expected returns with both historical observation and investors' views on asset performance, which in our case, the views from pairs trading strategy. The way we connect traditional pairs trading to the Black-Litterman portfolio optimization framework brings innovation to pairs trading strategy, and orchestras the methodology of statistical arbitrage and portfolio optimization.

In Section \ref{litreview}, we present the literature overview of methodology and empirical findings of Statistical Arbitrage and Black-Litterman model. So far very little literature focuses on establish a Black-Litterman model using statistical arbitrage strategy. In Section \ref{pairstrading}, we discuss how to select co-integrated stock pairs with rigorous Engle and Granger test. Then we transform the co-integration model into real world trading signals. Section \ref{portfolio} shows how to incorporate into the Black-Litterman model our prior views on stock returns based on pairs trading strategy. Finally, we calibrate model parameters and backtest the portfolio from two testing periods: out-of-sample 2016 to 2018 and stressful financial crisis in Section \ref{implement} We make our conclusions in Section \ref{conclusion}.

\section{Literature Overview}
\label{litreview}
\subsection{Statistical Arbitrage}
\subsubsection{Definition}
Statistical arbitrage (StatArb) is a trading strategy that seeks to exploit relative price movements between financial instruments. It leverages statistical and computational techniques to identify and capitalize on pricing inefficiencies. The approach involves constructing a portfolio of assets that are expected to converge to their historical price relationships, thus generating profit from this mean reversion behavior\citep{Avellaneda2010}.

The theoretical foundation of statistical arbitrage is rooted in the Efficient Market Hypothesis (EMH), which suggests that asset prices fully reflect all available information. StatArb strategies operate under the assumption that deviations from historical relationships are temporary and that prices will revert to their mean over time \citep{Gatev2006}. Key concepts of StatArb are:

\begin{itemize}
    \item \textbf{Mean Reversion:} The idea that asset prices will revert to their historical average over time \citep{Vidyamurthy2004}.
    \item \textbf{co-integration:} A statistical property of time series variables that indicates a long-term equilibrium relationship between them \citep{Elliott2005}.
    \item \textbf{Pair Trading:} A common form of StatArb where two correlated securities are traded against each other to exploit relative mispricings \citep{Pole2007}.
\end{itemize}

The more rigorous definition of a statistical arbitrage is a zero initial cost, self-financing trading strategy $x(t), t>0$ with cumulative discounted value $v(t)$ such that \citep{Hogan2004}

\begin{enumerate}
    \item \( v(0) = 0 \),
    \item \( \lim_{t \to \infty} \mathbb{E}_P[v(t)] > 0 \),
    \item \( \lim_{t \to \infty} \mathbb{P}(v(t) < 0) = 0 \),
    \item \( \lim_{t \to \infty} \text{Var}_P[v(t)] = 0 \) if \( \mathbb{P}(v(t) < 0) > 0 \) for all \( t < \infty \).
\end{enumerate}

In reality, most of the above conditions can be relaxed, and an investment portfolio can be described as a statistical arbitrage if it has the following features \citep{Elliott2005}

\begin{enumerate}
    \item trading signals are systematic
    \item the trading book is market neutral
    \item the mechanism for generating excess return is statistical
\end{enumerate}

\subsubsection{Implementations}
StatArb strategies utilize various statistical and econometric models to identify trading opportunities. Common methodologies include:

\begin{enumerate}
    \item \textbf{Pairs Trading:} Identifies pairs of stocks with historically correlated prices. When the price relationship diverges, the strategy involves shorting the over-performing stock and going long on the under-performing one, betting that they will converge \citep{Gatev2006}.
    
    \item \textbf{co-integration Analysis:} Uses econometric techniques to identify groups of assets with stable long-term relationships. The Johansen test is a popular method to determine co-integration among multiple assets \citep{Elliott2005}.
    
    \item \textbf{Machine Learning:} Recent advancements have seen the application of machine learning techniques such as neural networks, support vector machines, and clustering algorithms to improve the accuracy of predictions and enhance trading performance \citep{Avellaneda2010}.
    
    \item \textbf{Time Series Models:} Techniques like ARIMA (Auto-regressive Integrated Moving Average), GARCH (Generalized Auto-regressive Conditional Heteroskedasticity), and Kalman Filters are used to model and forecast price movements \citep{Vidyamurthy2004}.
\end{enumerate}

\subsubsection{Improvement}
Empirical research suggests that StatArb strategies tend to perform well in stable market conditions but may suffer during periods of high volatility or structural market changes \citep{Gatev2006}. Therefore, we propose in this research that we can improve on StatArb strategies with market observations or investor views under different market regime.

\subsection{Black-Litterman Framework}
The Black-Litterman model combines the equilibrium market returns with investor views to create a more stable and intuitive portfolio allocation \citep{Black1990}. It builds on the foundational work of mean-variance optimization. Mean-variance optimization is highly sensitive to the input parameters, the expected returns and covariance matrix. Research shows that robust covariance matrix with tuned parameters can mitigate the sensitivity to data noise \citep{Zhou2024}. The Black-Litterman model on the other hand addresses this issue of expected return by incorporating market equilibrium information and investor views in a Bayesian framework.

The model starts with the assumption that the market portfolio is in equilibrium, meaning the implied returns are proportional to the market capitalization of the assets. This equilibrium return vector, known as the "implied equilibrium returns", serves as a prior distribution in the Bayesian framework \citep{Black1990}.

Investors can incorporate their views on expected returns into the model. These views can be absolute (specific expected returns for certain assets) or relative (expected returns of one asset relative to another). The investor's views are represented mathematically and are combined with the equilibrium returns to form a posterior distribution of expected returns \citep{Idzorek2007}.

The Black-Litterman model uses a Bayesian approach to combine the equilibrium returns with investors' own views. The resulting posterior distribution provides a new set of expected returns, which are then used in the mean-variance portfolio optimization. This approach reduces the sensitivity to input parameters and provides more stable and realistic portfolio weights.

The process of applying the Black-Litterman model involves several key steps:

\begin{enumerate}
    \item \textbf{Determine Market Equilibrium Returns:} Calculate the implied equilibrium returns from the market capitalization weights and the covariance matrix of asset returns.
    \item \textbf{Specify Investor Views:} Formulate the investor's views on expected returns, including both absolute and relative views.
    \item \textbf{Combine Equilibrium Returns and Views:} Use the Black-Litterman formula to combine the equilibrium returns with the investor's views, resulting in a new set of expected returns.
    \item \textbf{Optimize Portfolio:} Apply mean-variance optimization using the adjusted expected returns and the covariance matrix to determine the optimal portfolio weights.
\end{enumerate}

\section{Pairs Trading}

\label{pairstrading}

Pairs trading is a widely adopted implementation as mentioned above and takes up to 40\% of Statistical Arbitrage strategies in the market. The strategy is based on the idea that market investors often overreact or under-react to new information, causing prices to oscillate and deviate from its pair value. A contrarian approach can profit if stock prices are to eventually return to their equilibrium. To create a statistical arbitrage opportunity, the paired stocks should exhibit similar historical price movements and correlated returns.

To form a pairs trading, we performed co-integration analysis on the returns of two stocks, $S1$ and $S2$. Among various pairs selection methods, such as the distance method, stationarity criterion, and co-integration, the co-integration modeling provides the highest and the most persistent returns across different economic regime \cite{Huck2015}. The co-integrated return system is modeled by the following stochastic differential equation:

\[
\frac{dS_1(t)}{S_1(t)} = a \, dt + b \, \frac{dS_2(t)}{S_2(t)} + dX(t)
\]

$X(t)$ is a stationary and mean-reverting process. $a$ is the drift parameter that can be ignored because its value is small over a short time period \cite{Avellaneda2010}. Therefore, the strategy should long $1$ dollar of stock $S_1$ and short $b$ dollars of stock $S_2$ when $X(t)$ is small and tends to increase to its long-term equilibrium. Similarly, it will short $1$ dollar of stock $S_1$ and long $b$ dollars of stock $S_2$ when $X(t)$ is large and tends to decrease to the equilibrium. There will be a positive return if the model accurately captures the return co-integration of stocks $S{1}$ and $S{2}$. 

There are three steps in designing the experimental strategy: 1. prescreen stock universe, 2. perform co-integration test and select co-integrated stock pairs 3. formulate a realistic trading signal. We will explain each step in the following sections.


\subsection{Prescreening Approach}
\label{prescreen}
It is impractical to run co-integration tests on every possible stock pair in the U.S stock market. For instance, the S\&P 500 alone comprises 505 common stocks will lead to a staggering 127,260 pairs. Therefore, we employ a prescreening approach \cite{Vidyamurthy2011} to identify stocks that are most likely to exhibit significant co-movement with other stocks. Candidate stock pairs are selected based the following criteria: sector and size (i.e. market capitalization). Since stock returns are largely driven by common risk factors, we believe stocks in the same sector and stocks of similar size with should be more likely to have co-integrated returns. In our approach, we consider top 5 holdings from each sector of SPY ETF and the resulting stock universe includes 45 stocks. With the presceening, we are left with 90 pairs of stocks to be examined at each portfolio rebalance date. It significantly improves the efficiency of stock pair selection.

\subsection{Co-integration Test}
\label{co-integration test}
We analyzed the co-integration by performing a Engle and Granger (EG) test on discretized return time series within a predetermined time window \( w \) to  \cite{Engle1987}:
\begin{align*}
    n &= a + bR_{2n} + e_n, \quad n = 1, 2, \ldots, w \\
\end{align*}

With \( R_{1n} \), \( R_{2n} \), and \( e_n \) being the discretization of \( \frac{dS_{1t}}{S_{1t}} \), \( \frac{dS_{2t}}{S_{2t}} \), and \( dX_t \). We also define the cumulative residual process \( X_n = \sum_{k=1}^{n} e_k \), \( n = 1, 2, \ldots, w \), as the discretized time series of \( X_t \). co-integration \( S_{t1} \) and \( S_{t2} \) depends on the stationarity of the discretized cumulative residual process \( X_n \), which is tested by the Augmented Dickey-Fuller test (ADF). The process is outlined as below:
\begin{enumerate}
    \item Run an OLS regression on \( R_{1n} \) on \( R_{2n} \)
    \item Obtain \( \hat{a} \), \( \hat{b} \), and residual \( \hat{e}_n \)
    \item Calculate cumulative residual: \( \hat{X}_n = \sum_{k=1}^{n} \hat{e}_k \) for \( n = 1, 2, \ldots, w \)
    \item Run the Augmented Dickey-Fuller test on \( \hat{X}_n \) to test the stationarity
\end{enumerate}

This co-integration test is performed on each window of historical data. If stationarity is rejected, then the pair of stocks should not be considered. We chose the window size \( w \) to be 60 days as it is approximately the length of one earnings cycle.

\begin{figure}
	\centering
        \includegraphics[width=0.5\textwidth]{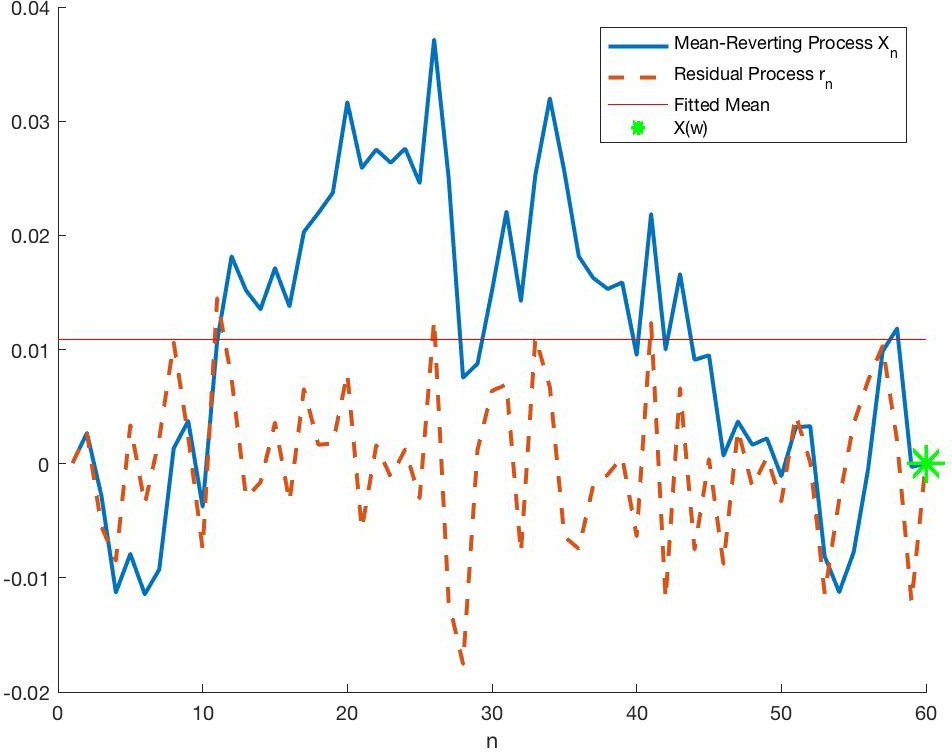} 
	\caption{Fitted OU process with window size 60}
	\label{fig3}
\end{figure}

After we validates the co-integration relationship, we fit the residual process \( X(t) \) with Ornstein-Uhlenbeck (OU) stochastic process:

\[
dX(t) = k(m - X(t)) \, dt + s \, dW(t); \quad k > 0
\]

\( k \) is the parameter that determines the strength of mean-reversion, and \( m \) is the equilibrium mean of this stochastic process. Figure \ref{fig3} shows a calibrated OU process with time window \( w = 60 \). In this example, We chose 9 top holding stocks in each of 9 sectors in SPY. Return is obtained by daily adjusted close price. The figure shows the mean-reverting process $X_n$, residual process, the fitted mean and the fitted \( X(w) = X(60) = 0 \) at the green asterisk.

The parameter \( k \) in the OU process can be interpreted as the speed of mean-reversion, and \( t = \frac{1}{k} \) can is the time scale of mean reversion. We select pairs with the fastest mean-reversion, ideally with \( t << w \) because they have more arbitrage opportunities.

\subsection{Trading Signal Formulation}
\label{sec1.4}
The OU process is a stationary Gaussian process and it means that every finite collection of \( X(t) \) indexed by time has a multivariate normal distribution. Therefore, \( X(t) \) can be considered as a normal random variable. To construct the trading signal, we consider normalizing \( X(t) \). The equilibrium variance of the OU process is:

\[
s_{\text{eq}} = \frac{s}{\sqrt{2k}}
\]

We can then define the normalized dimensionless variable based on the variance as our trading signal:

\[
s = \frac{X(t) - m}{s_{\text{eq}}}
\]

This \( s \)-score indicates how far away the return spread of this pair of stocks is from the theoretical equilibrium. Note that for each regression in a 60-day time window, fitting the intercept forces \( X(60) = 0 \). Therefore, the \( s \)-score on the last day becomes:

\[
s = -\frac{m}{s_{\text{eq}}}
\]

The following decisions are made based on the value of this score:

\begin{itemize}
    \item Buy to open: long 1 dollar of stock \( S_1 \), and short \( \hat{b} \) dollar of stock \( S_2 \) if \( s < -s_{\text{bo}} \)
    \item Sell to open: short 1 dollar of stock \( S_1 \), and long \( \hat{b} \) dollar of stock \( S_2 \) if \( s > +s_{\text{so}} \)
    \item Close short position if \( s < +s_{\text{bc}} \)
    \item Close long position if \( s > -s_{\text{c}} \)
\end{itemize}

Intuitively, \( s_{\text{bo}} \) and \( s_{\text{so}} \) should be much larger than \( s_{\text{bc}} \) and \( s_{\text{c}} \), as we aim to enter the position when \( X(t) \) is far away from equilibrium. On the other hand, when \( X(t) \) has converged near the equilibrium, it is uncertain whether the previous trend will continue, and thus we should close the position.

\section{Portfolio Optimization with Black-Litterman}
\label{portfolio}
\subsection{Posterior Return Estimation}

We established the following model to incorporate the pairs trading strategy into the Black-Litterman model. For return \( \mathbf{m} \):

\[
p(\mathbf{m}|\mathbf{X}) \sim L(\mathbf{X}|\mathbf{m}) \cdot g(\mathbf{m})
\]

\( \mathbf{X} \) represents the observed data or historical stock returns. \( p(\mathbf{m}|\mathbf{X}) \) is the posterior distribution of the parameters, \( L(\mathbf{X}|\mathbf{m}) \) is the likelihood function for the observed data, and \( g(\mathbf{m}) \) is the prior view on the distributions of return parameters.

Assuming there are \( N \) assets in the market with returns distributed normally with mean \( \mathbf{m} \) and covariance matrix \( \mathbf{S} \), denoted as \( R \sim N(\mathbf{m}, \mathbf{S}) \). Assuming \( R_1, R_2, \ldots, R_T \) are independent and identically distributed, we have prior estimation:

\[
\hat{\mathbf{m}} = \frac{1}{T} \sum_{t=1}^{T} \mathbf{r}_t \sim N\left(\mathbf{m}, \frac{1}{T} \mathbf{S}\right)
\]

We express our views on the expected returns using a linear system of equations. Let \( K \) be the total number of views, \( \mathbf{P} \) be a \( K \times N \) picking matrix, \( \mathbf{Q} \) be a \( K \)-dimensional view vector. Each row in \( \mathbf{P} \) expresses a view of all assets.


We assume that \( \mathbf{Q} \sim N(\mathbf{q}, \mathbf{W}) \), where \( \mathbf{q} \) is the expected returns according to the views, and \( \mathbf{W} \) is the diagonal covariance matrix of the expressed views, representing the uncertainty in each view. The posterior return \(\bar{\mathbf{m}}\) and covariance matrix \( \mathbf{M} \) of the posterior distribution are:

\[
\bar{\mathbf{m}} = \left(\mathbf{S}^{-1} + \mathbf{P}^T \mathbf{W}^{-1} \mathbf{P}\right)^{-1} \left(\mathbf{S}^{-1}\hat{\mathbf{m}}^T + \mathbf{P}^T \mathbf{W}^{-1} \mathbf{q}\right)
\]

\[
\mathbf{M} = \left(\mathbf{S}^{-1} + \mathbf{P}^T \mathbf{W}^{-1} \mathbf{P}\right)^{-1}
\]


\subsection{Model Formulation}
We constructed the picking matrix \( \mathbf{P} \) and the distribution of the random vector \( \mathbf{Q} \). Suppose there are \( N \) assets and \( K \) pairs to be expressed as the prior, the matrix \( \mathbf{P} \) will have size \( K \times N \). We have matrix \( \mathbf{P} \) of the following form:
\[
\mathbf{P} =
\begin{pmatrix}
1 & -b_1 & 0 & 0 & \cdots & 0 & 0 \\
0 & 0 & 1 & -b_2 & \cdots & 0 & 0 \\
\vdots & \vdots & \vdots & \vdots & \ddots & \vdots & \vdots \\
0 & 0 & 0 & 0 & \cdots & 1 & -b_K \\
\end{pmatrix}
\]
Each row includes an co-integrated stock pair, \( b_i \) for \( i = 1, \ldots, K \) are the fitted slopes of OLS and \( k_i \), \( m_i \), and \( s_{eq,i} \) for \( i = 1, \ldots, N \) are fitted coefficients in the OU process in Section \ref{co-integration test}. \( \mathbf{q} \) is the expected return of each pairs trading strategy, and \( \mathbf{W} \) the covariance matrix of their returns. A constant \( l \) is multiplied on each return in \( \mathbf{q} \) that indicates the how convicted the beliefs are:
\[
\mathbf{q} =
\begin{pmatrix}
lk_1(m_1 - X_1(t)) \Delta t \\
lk_2(m_2 - X_2(t)) \Delta t \\
\vdots \\
lk_K(m_K - X_K(t)) \Delta t \\
\end{pmatrix}
\]
\[
\mathbf{W} =
\begin{pmatrix}
s^2_{eq,1} \Delta t & 0 & \cdots & 0 \\
0 & s^2_{eq,2} \Delta t & \cdots & 0 \\
\vdots & \vdots & \ddots & \vdots \\
0 & 0 & \cdots & s^2_{eq,K} \Delta t \\
\end{pmatrix}
\]

\subsection{Portfolio Optimization Formulation}
we considered market frictions cost and experimented with leverage constraints in the portfolio optimization.

\begin{itemize}
    \item Market friction cost: In the case of forming a statistical arbitrage portfolio, transaction costs and short-sale constraints are the two most non-negligible expenses that will make a visible impact on the returns. We assume a fixed transaction cost of $5$ bps per trade executed, for both opening or closing of a long or short position. The total transaction cost is then subtracted from the profit on a daily basis. A short-sale constraint is incorporated into the implementation such that an amount proportional to the value of the short asset is deducted daily from the profit throughout the short period. The annualized short-sale cost is assumed to be 1\%.
    
    \item Portfolio Rebalance: The portfolio is rebalanced based on the pair selection frequency: 60 days. This is a hyperparameter that we tuned in Section \ref{calibration}. In the pair-selection process, we liquidate holdings of the previous stock pairs. The total wealth is then distributed equally into sectors where a valid stock pair was chosen by the co-integration test.
    
    \item Leverage: Since pairs trading usually has much smaller return volatility than that of a normal trading strategy, we decide to add leverage to amplify the return while maintaining the volatility around a target level. That is, the profit/loss is multiplied by the leverage ratio less the cost incurred in borrowing on a daily basis. A fixed annualized borrowing rate of 2\% is used in the implementation; however, in reality, the rate could be even lower through negotiation if one has a significant amount of capital.
    
    \item The returns are calculated based on the daily cumulative wealth of the portfolio. The end-of-day wealth in a sector with an actively traded pair is computed as the initial wealth plus the profit (or loss) in the long/short position less the leverage and short-sale costs. In a sector where a valid pair has been selected but the trading (open) signal is not present, the wealth is invested in the money market and grows at the assumed annualized risk-free interest rate of 2\% or invested in the S\&P 500 index. The sum of wealth in all sectors at the end of the day is thus the cumulative wealth of the portfolio.
\end{itemize}

We input the expected return \( \bar{\mathbf{m}} \) to the optimization: 
\[\text{max } \bar{\mathbf{m}}^T \mathbf{x} - \frac{1}{2} d\mathbf{x}^T \mathbf{Sx}\]
Here, \( \frac{1}{2}d\mathbf{x}^T \mathbf{Sx} \) can be interpreted as the risk penalty, and \( d \) is the user-defined parameter for risk-aversion.

The optimization constraints include:
\begin{itemize}
    \item the sum of weights in each stock added to the total transaction cost equals $1$;
    \item the per-trade transaction cost times the sum of change in positions must not exceed a limit for the total transaction cost;
    \item the weighted position in any stock is bounded by $-1$ and $1$;
    \item the total per-unit transaction cost must be less than $0.02$.
\end{itemize}

\section{Implementation and Performance Analysis}
\label{implement}

\subsection{Data and Model Implementation}
\label{sec1.5}
We used 2006 to 2016 as the training set and 2016 to 2018 as the out-of-sample test set. All parameters in Section \ref{calibration} is calibrated with training set in a time series cross validation fashion, which avoids look-forward bias. 

We conducted presreening to construct a small stock universe as explained in Section \ref{prescreen}, then performed co-integration test on each stock pair. Our final portfolio is composed of 9 pairs of stocks from 9 sectors, totaling 18 stocks. Daily stock returns are calculated based on the adjusted closing prices. 

If a co-integrated stock pair triggers open signal, we will add a new row to the matrices \( \mathbf{P} \), \( \mathbf{q} \), and \( \mathbf{W} \), expressing our view that the pairs will likely generate positive returns in the future. The posterior return \( \bar{\mathbf{m}} \) along with the original covariance matrix \( \mathbf{S} \) are fed into the mean-variance optimization problem to solve for an optimal weight vector \( \mathbf{x} \). We chose the risk aversion parameter \( d \) to be 2.


\subsection{Parameter Calibration}
\label{calibration}
We tuned three key parameters: the mean reversion speed, frequency of stock pair selection, cutoffs for the s-score.

First, a fast mean reversion speed would require executing multiple trades in a short time period and may incur high transaction costs, while a slow mean reversion speed would cause the problem of holding the positions for too long and positions will be forced to close early by the next pair-selection process. There could be loss because we cannot capture the appropriate closing signal. In our strategy, we consider pairs that have a mean reversion period of at least one trading week (5 trading days) and it shoud be less than the frequency of pair-selection. Among all the qualified pairs in a sector, we chose the pair with the fastest mean reversion speed.

Another parameter we need to tune is the frequency of stock pair selection. The profitability of the strategy is highly dependent on the choice of frequency. Since the mean reversion speed is at least 5 trading days, the frequency should be greater than 10 days to ensure that the profit generated by the pairs can be fully captured. On the other hand, a small selection frequency indicates a strong conviction that the same mean reversion trend will persist for a long time in a market, counter-intuitive to empirical facts. Therefore, we calibrate the selection frequency based on the Sharpe ratio and the Calmar ratio of the portfolio. The Sharpe ratio is a measure of the return in excess of the risk-free rate per unit of volatility. The Calmar ratio, computed as the return over the maximum drawdown, measures the performance of a portfolio relative to its risk. Upon testing on a set of rational frequency choices on the non-leveraged strategy, as shown in Table \ref{pair-selection}, we found a selection frequency of $27$ trading days produced both the highest Sharpe ratio and the Calmar ratio in the training period.

Finally, we calibrates cutoffs for the s-score based on the frequency parameter and we chose \( s_{bo} = 1.3 \), \( s_{so} = 1.3 \), \( s_{bc} = 0.7 \), and \( s_{sc} = 0.5 \).

\begin{table}[htbp]
    \centering
    \caption{Pair-selection Frequency Tuning}
    \label{pair-selection}
    \begin{tabular}{ccccccc}
        \hline
        Freq. & Sharpe & Calmar & Freq. & Sharpe & Calmar & Freq. \\
        \hline
        11 & 0.0939 & 2.1402 & 18 & 1.3469 & 3.6605 & 25 \\
        12 & 0.9177 & 2.8157 & 19 & -0.9014 & -0.3123 & 26 \\
        13 & 1.6258 & 3.7281 & 20 & 0.7164 & 2.3237 & 27 \\
        14 & -0.9251 & -0.0883 & 21 & -2.3884 & -1.1930 & 28 \\
        15 & 0.0266 & 0.7829 & 22 & -0.0784 & 0.7926 & 29 \\
        16 & 0.6446 & 1.9423 & 23 & 0.3534 & 1.3331 & 30 \\
        17 & 0.0580 & 0.8743 & 24 & 0.7690 & 2.7464 & \\
        \hline
    \end{tabular}
\end{table}

\subsection{Performance Evaluation}
We implemented an enhanced index portfolio "Pairs Trading + S\&P 500 Index". It trades the identified stock pairs when the trading signals are triggered, and trading decisions will be optimized into an efficient portfolio. We long the S\&P 500 index when there are no signals. This portfolio behaves as an enhanced index strategy because it is closely correlated with the market return while exhibiting superior return.

The portfolio was backtested from 2016 to 2018. Table \ref{strategy} compares between the performance of the index enhanced portfolio and the S\&P 500 index. We can see that the strategy outperforms the S\&P 500 index in all of the performance measure even without leverage. It also has notably lower volatility and less max drawdown than S\&P 500 index.

\begin{figure}
	\centering
        \includegraphics[width=0.7\textwidth]{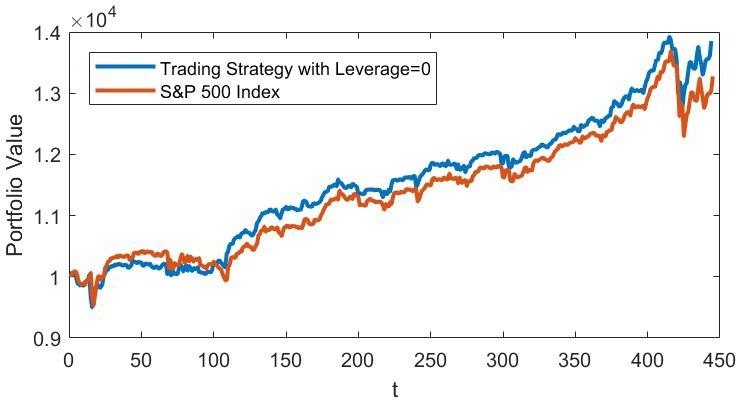} 
	\caption{Out-of-sample Portfolio Cumulative Growth}
	\label{strategy_fig}
\end{figure}

From Figure \ref{strategy_fig}, we can see that our portfolio consistently outperforms the S\&P 500 index and suffers less in extreme market flash crashes. Additionally, the strategy also has a strong return correlation with S\&P index due to the fact that it invests in S\&P index when no pairs trading signal is triggered.

\begin{table}[htbp]
    \centering
    \caption{Out-of-sample Performance statistics without Leverage}
    \label{strategy}
    \begin{tabular}{lccccc}
        \hline
        Strategy & Total Return & Average Return & Sharpe & Volatility & Max Drawdown \\
        \hline
        S\&P 500 & 0.3275 & 0.1636 & 1.4081 & 0.1020 & 0.1016 \\
        Leverage=0 & 0.3785 & 0.1871 & 1.7808 & 0.0938 & 0.0844 \\
        \hline
    \end{tabular}
\end{table}

\begin{table}[htbp]
    \centering
    \caption{Out-of-sample Performance statistics with 3x Leverage}
    \label{tab:performance-comparison}
    \begin{tabular}{cccccc}
        \hline
        & Total Return & Average Return & Sharpe & Volatility & Max Drawdown \\
        \hline
        Leverage=0 & 0.1048 & 0.0569 & 1.6837 & 0.0219 & 0.0085 \\
        S\&P 500 & 0.3275 & 0.1636 & 1.4081 & 0.1020 & 0.1016 \\
        Leverage=3.65 & 0.4611 & 0.2209 & 1.9711 & 0.1019 & 0.0430 \\
        \hline
    \end{tabular}
\end{table}

\subsection{Stress Testing}
\begin{figure}
	\centering
        \includegraphics[width=0.7\textwidth]{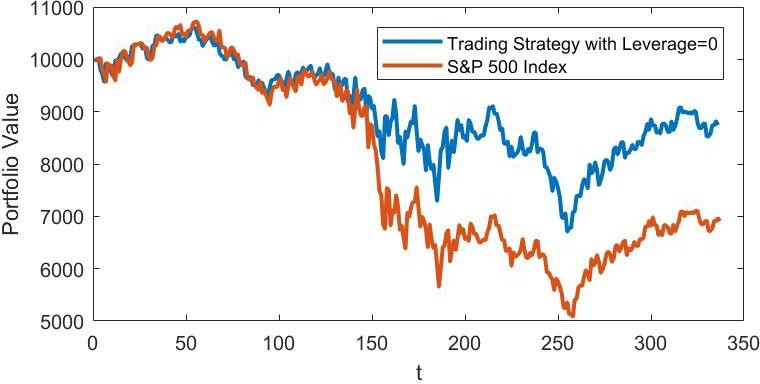} 
	\caption{Financial Crisis: Portfolio Cumulative Growth}
	\label{strategy_fig_crisis}
\end{figure}

To test the robustness, we also constructed portfolios during the financial crisis from Dec 2007 to June 2009. Figure \ref{strategy_fig_crisis} and Table \ref{strategy_table_crisis} shows the portfolio value of our strategy. The return exhibits a similar pattern to the S\&P 500 index, especially in the first 100 days. However, our portfolio has a higher return over the S\&P 500 during the later period.

\begin{table}[htbp]
    \centering
    \caption{Financial Crisis: Performance statistics}
    \label{strategy_table_crisis}
    \begin{tabular}{lccccc}
        \hline
        Strategy & Total Return & Average Return & Sharpe & Volatility & Max Drawdown \\
        \hline
        S\&P 500 & -0.3091 & -0.1957 & -0.5294 & 0.4074 & 0.5258 \\
        Leverage=0 & -0.1262 & -0.0416 & -0.1774 & 0.3470 & 0.3679 \\
        \hline
    \end{tabular}
\end{table}

\section{conclusion}
\label{conclusion}
We first discussed how to implement pairs trading through co-integration modeling and demonstrated that cointegrated stock pairs can  generate excess returns with small risk. Then we incorporated pairs trading as a return prior into the Black-Litterman portfolio optimization framework. The out-of-sample backtest result proves that our strategy can outperform the S\&P500 index and suffered less drawdown during financial crisis period. It has great implication on how we can manage the risk of traditional pairs trading strategy in a more systematic manner.

\bibliographystyle {unsrtnat}






\end{document}